\documentclass[manuscript,screen]{acmart}

\AtBeginDocument{%
  \providecommand\BibTeX{{%
    \normalfont B\kern-0.5em{\scshape i\kern-0.25em b}\kern-0.8em\TeX}}}

\copyrightyear{2020} 
\acmYear{2020} 
\setcopyright{acmcopyright}\acmConference[PEARC '20]{Practice and Experience in Advanced Research Computing}{July 26--30, 2020}{Portland, OR, USA}
\acmBooktitle{Practice and Experience in Advanced Research Computing (PEARC '20), July 26--30, 2020, Portland, OR, USA}
\acmPrice{15.00}
\acmDOI{10.1145/3311790.3400848}
\acmISBN{978-1-4503-6689-2/20/07}

\usepackage{balance}

\begin{document}

\title{Toward Interoperable Cyberinfrastructure: Common Descriptions for Computational Resources and Applications}

\author{Joe Stubbs}
\affiliation{%
  \institution{TACC, University of Texas, Austin}}
\email{jstubbs@tacc.utexas.edu}

\author{Suresh Marru}
\affiliation{%
  \institution{Cyberinfrastructure Integration Research Center, Indiana University}}
\email{smarru@iu.edu}

\author{Daniel Mejia}
\affiliation{%
  \institution{Network for Computational Nanotechnology, Purdue University}}
\email{dmejiapa@purdue.edu}

\author{Daniel S. Katz}
\affiliation{%
  \institution{NCSA, CS, ECE, iSchool, University of Illinois at Urbana-Champaign}}
\email{d.katz@ieee.org}

\author{Kyle Chard}
\affiliation{%
  \institution{University of Chicago}}
\email{chard@uchicago.edu}

\author{Maytal Dahan}
\affiliation{%
  \institution{TACC, University of Texas, Austin}}
\email{maytal@tacc.utexas.edu}

\author{Marlon Pierce}
\affiliation{%
  \institution{Cyberinfrastructure Integration Research Center, Indiana University}}
\email{marpierc@iu.edu}

\author{Michael Zentner}
\affiliation{%
  \institution{San Diego Supercomputer Center, University of California, San Diego}}
\email{mzentner@ucsd.edu}

\renewcommand{\shortauthors}{Stubbs and Marru, et al.}
\renewcommand{\shorttitle}{Common Descriptions for Computational Resources and Applications}
\begin{abstract}
The user-facing components of the Cyberinfrastructure (CI) ecosystem, science gateways and scientific workflow systems, share a common need of interfacing with physical resources (storage systems and execution environments) to manage data and execute codes (applications). However, there is no uniform, platform-independent way to describe either the resources or the applications. To address this, we propose uniform semantics for describing resources and applications that will be relevant to a diverse set of stakeholders. We sketch a solution to the problem of a common description and catalog of  resources: we describe an approach to implementing a resource registry for use by the community and discuss potential approaches to some long-term challenges. We conclude by looking ahead to the application description language.

\end{abstract}


\keywords{Cyberinfrastructure, interoperability, resource description, application description, science gateways, science gateways community institute}

\maketitle

\section{Introduction}
A wide range of cyberinfrastructure (CI) components exist within the overall scientific ecosystem, from science gateways to scientific workflow systems to Jupyter notebook-based environments, that integrate and provide access to scientific applications, computing resources, and scientific data collections. For example, the Science Gateways Community Institute’s~\cite{sgci-survey, sgci-2yrs} gateway catalog~\cite{sgci-catalog} alone contains nearly 600 entries, and Globus~\cite{globus-saas} provides access to 20,000 active endpoints. Typically, these systems rely on a variety of storage and execution infrastructure to maintain data collections and perform analyses. Such resources range from departmental servers and campus clusters, to regional and national scale supercomputing facilities, to academic and commercial clouds, as well as the unique communities and applications themselves. There is presently no uniform way of describing and discovering these resources and applications, and as a result, each CI project creates its own ways of encoding this information. 

This non-standardized approach leads to two problems. First, the information must be discovered and recompiled by each cyberinfrastructure component as part of configuring and deploying the component, and as a result, the information is often incomplete or inaccurate. Second, the components themselves, such as data management tools or analysis components, are coupled to the non-standard description of the resources, typically through configuration metadata, and as a consequence, there is a barrier to reusing the component in another framework or platform. 

To address this problem, we propose a  roadmap for improving interoperability across CI components, and we sketch a solution to the challenge of resource description and discovery: a centralized, uniform computational resource registry. This would be valuable in its own right: CI components would be able to automatically obtain complete and up-to-date information about available resources, removing the laborious and error-prone manual process currently in place. In turn, the resource registry could improve aggregate resource utilization and reduce time-to-solution. 

While we ultimately strive for interoperability at the scientific application level, finding a solution for uniform resources is a necessary precursor. Codes for computational research tend to be highly sensitive to the execution environment, often depending on specific hardware architectures, software libraries, storage resources, etc. to run correctly, much less optimally. 

\section{Stakeholders and Prior Work}

A number of common CI frameworks and components leverage custom, non-standard ways of describing resources and/or applications. The extensive experience of these stakeholders provides a starting point for understanding requirements and defining initial solutions. 

The \textbf{Tapis} framework~\cite{tapis} provides APIs to enable gateway developers and researchers to automate interactions with high-performance computing (HPC) and cloud resources. Fundamental to the Tapis system are its Systems and Apps services: users define the resources (systems) and analysis codes (apps) their programs will use by making HTTP requests to the Systems and Apps APIs, respectively. 

\textbf{Apache Airavata}~\cite{airavata} is a science gateway framework, hosted and operated as a  platform instance that connects many gateways to diverse computing and storage resources. It provides abstractions for managing the execution and provenance of applications on diverse computational resources ~\cite{scigap-ar-2018}.   
\textbf{HUBzero}~\cite{hubzero} is an open source software platform on which to create science gateways.  The platform provides capabilities to host scientific tools, publish data, share resources and build communities. HUBzero-based science gateways combine web interfaces with a middleware that provides access to interactive simulation tools running on cloud computing resources, campus clusters, and other national HPC facilities.  

\textbf{Globus}~\cite{globus-saas} is a high-performance data management service and platform that is used across many  CI projects and frameworks, including Tapis, Airavata, HubZero, Whole Tale, and Parsl. Globus endpoints can be deployed on laptops, supercomputers, tape archives, scientific instruments and cloud storage.

\textbf{Parsl}~\cite{babuji19parsl, 10.1145/3332186.3332231} is a Python parallel scripting library that allows Python programs to be written with task control and dependency logic separated from resource configuration information. Parsl constructs a dynamic dependency graph of components that it can then execute efficiently on diverse resources.

The \textbf{WholeTale}~\cite{BRINCKMAN2019854} platform is an open, web-based service that allows researchers to create “tales” -- reproducible artifacts that combine data, computation environment, and the narrative of computational research (e.g., in a Jupyter notebook).  

\section{Uniform Resource Descriptions: A Solution Sketch}
\textbf{Resource Description Specification:} To define the resource description specification, we begin with an overview of the technical scope. The goal is to include attributes into the specification that allow for describing as many CI resources as possible. Table~\ref{tab:res-types} summarizes the types of resources considered to be in-scope for this specification.
In general, the resource is required to be available on a network accessible by other machines, though it need not be on a public network. Additionally, Table~\ref{tab:resource-metadata-types} describes the metadata that are in-scope.

\begin{table}
  \caption{Resource Types}
  \label{tab:res-types}
    \begin{tabular}{p{0.30\columnwidth}p{0.65\columnwidth}}
    \toprule
    Category & Examples \\
    \midrule
    HPC and campus compute clusters %
&
    Blue Waters, Bridges, Comet, Frontera, OSC cluster, Stampede2 %
\\
    National-scale storage systems %
& 
        Corral, Globus Endpoints, SDSC Openstack Storage, Stockyard %
\\
    Academic or Commercial Clouds %
 &
    JetStream or EC2 instances, S3 Buckets %
\\
   Individual or Lab Resources %
& 
   Laptops, Workstations, Compute clusters (e.g., Hadoop, Spark, etc) %
\\
    \bottomrule
    \end{tabular}
\end{table}

\begin{table}
  \caption{Types of resource metadata}
  \label{tab:resource-metadata-types}
    \begin{tabular}{lp{0.65\columnwidth}}
        \toprule
Category & Description \\
    \midrule
High-level data & 
Primary information about the resource, including hostname or network address, owner, and type (e.g., “compute” or “storage”)%
\\
Hardware & 
Details about the hardware, including CPU architecture, memory type, core and thread count, storage type and driver, network type, etc.%
\\
Operating System & 
Information about the operating system, including kernel and distribution versions%
\\
Scheduler &
Information regarding the scheduler, including scheduler type (e.g., “slurm” or “sge” but also “fork” for resources not using batch scheduling) and version, queue definitions, etc.%
\\
Software & 
Information about available software packages, including MPI/OpenMP, CUDA, container runtimes, etc.%
\\
    \bottomrule
    \end{tabular}
\end{table}

\textbf{Design and Implementation of Centralized Resource Registry:} The design and implementation of the centralized resource registry must meet several requirements. 1) The registry must contain accurate, up-to-date information, as the goal is to have production CI systems depending on this information. 2) The system must make it easy to add new resources, modify, and maintain existing information as changes are made to the underlying resources (e.g., additional nodes added to a queue on a resource). 3) The registry itself must be highly available because this impacts the availability of the CI tools using it. 4) Finally, the physical cost of building and maintaining the registry must be kept to a minimum, as this is critical to its sustainability.

Building the registry on top of a highly-available, cloud-hosted service helps meet the cost and availability goals. Storing registry information, including the versions of the specification, across one or more repositories hosted in a service such as GitHub is an attractive bootstrapping option. In addition to providing free, highly available hosting, GitHub provides issue tracking and fine-grained authorization controls with familiar workflows such as pull requests. The GitHub Pages feature can host web-accessible applications with minimal effort, including dynamic applications written with JavaScript. To insulate downstream systems from changes, both resource definitions and the resource specification itself should be versioned. Consuming systems will be able query the repository for a specific version of a resource or for the “latest” version. 

Over time, we envision complementing the registry with additional tools, as defined by the needs of the community. For example, a tool for automatically validating a resource definition could help reduce errors and the time required by humans to maintain the definitions. Additionally, more robust search capabilities will facilitate automated discovery of resources.In providing new tools, we will leverage existing solutions wherever possible to minimize maintenance cost. For example, schema validation can be provided by using JSON Schema to define the specification itself and then leveraging the reference implementations of tools such as validators. 

Other issues must be addressed in future steps. For example, tens to hundreds of thousands of resources will be added to the registry over time, considering the growth of edge computing. With this growth, performance and scalability of the registry could become an issue. We must also consider resource lifecycles: stale entries in the registry (for old resources) need to be retained for archival purposes, so decluttering the registry and other maintenance will be needed to keep the information reliable.

\textbf{Adoption in CI Components:} The final step in implementing uniform resource descriptions is fostering adoption across CI components. The value proposition of the project rests on improving interoperability across CI components. If a critical mass of CI projects adopt the usage of uniform resource descriptions, 1) developers and users of the participating components will be able to discover accurate information about resources; and 2) participating CI projects will be invested in the maintenance and success of the registry.

With support from the Science Gateways Community Institute (SGCI), we will promote our project with leading CI component developers with the hopes of obtaining commitments to support and implement the specification and registry. We plan to organize BoF sessions and workshops at conferences, including Gateways, WORKS, and SCXY, to solicit input and participation from the CI community. We will use channels available to us through SGCI, such as the webinar series, for dissemination of progress. Finally, once we establish an initial public project space, such as a GitHub repository, we will enable community input via its associated mechanisms, such as its issue tracker.

\section{Uniform Application Descriptions}
We begin building a uniform application description by defining the types of applications in scope (in Table~\ref{tab:app-types}), followed by the types of metadata that need to be included (in Table~\ref{tab:app-metadata-types}). The situation is less clear cut than for resources, and the definitions of application types will need further discussion. 

\begin{table}
  \caption{Application  types}
  \label{tab:app-types}
    \begin{tabular}{lp{0.65\columnwidth}}
    \toprule
    Type & Examples \\
    \midrule
Command-line/batch & 
bioinformatics (fastqc, RNAseq), HPC (AMBER, LS-DYNA), engineering (opensees, ADCIRC), data vis (paraview), utilities (compression, format converters)  %
\\
Interactive & 
Jupyter, Matlab, RStudio %
\\
Streaming & 
Spark Streaming, Apache Storm %
\\
    \bottomrule
    \end{tabular}
\end{table}

\begin{table}
  \caption{Types of application metadata}
  \label{tab:app-metadata-types}
    \begin{tabular}{p{0.30\columnwidth}p{0.65\columnwidth}}
    \toprule
    Category & Description \\
    \midrule
High-level data %
& 
Primary information about the application, including application name, type, and description %
\\
Packaging %
 &
Data pertaining to the application packaging, including container or VM image, unikernel, module, etc. %
\\
Architecture \& Hardware Dependencies %
 &
Architecture (SP/MP) and hardware dependencies (processors, storage, networks, etc) %
\\
Software Dependencies %
&
Frameworks (Hadoop), Schedulers (SLURM), libraries, modules, etc. %
\\
Inputs and Data Dependencies %
 &
Input files, objects, databases, URLs, environment variables %
\\
Runtime Requirements %
 &
Additional runtime requirements, e.g., run as UID/GID, OS capabilities %
\\
Outputs %
&
Products produced by the app, including files, stdout streams, etc. %
\\
    \bottomrule
    \end{tabular}
\end{table}

\section{Related Work}
DRMAA~\cite{drmaa.v2} defined  a  generalized  API  to distributed  resource  management  systems  to  facilitate  the development of portable application programs and high-level libraries. JSDL~\cite{jsdl.56} describes the requirements of computational jobs for submission to resources in grid environments. GLUE~\cite{glue.v2} is  a conceptual information  model  for  grid  entities  described  using  the  natural  language  and  UML  class  diagrams. Globus Resource Specification Language~\cite{globus1998} described grid resources including computational job information. The Simple API for Grid Applications (SAGA)~\cite{saga} provides high-level interfaces for common grid components (e.g., transfer and scheduling). The Globus  Monitoring and Discovery Service \cite{globus-mds} built upon the LDAP protocol to address the distributed resource selection problem.

\section{Conclusion}
This paper discusses early stage efforts in working towards a common computational resource schema. This is being bootstrapped by cyberinfrastructure practitioners committed to adopt the resulting common descriptions into respective projects. The next phase will develop the schema and reference implementations. We plan to engage the cyberinfrastructure developer community as broader adoption of the proposed common language will facilitate interoperability amongst various  cyberinfrastructure components. 
\begin{acks}
JS, SM, MZ, MD and MP are supported by NSF award \#1547611. DSK and KC are supported by NSF award \#1550588. KC is also supported by NSF award \#1541450.
\end{acks}
\balance

\bibliographystyle{ACM-Reference-Format}
\bibliography{bib-pearc20-common-resource-description}

\end{document}